\documentclass[aps,pre,twocolumn,showpacs,superscriptaddress,groupedaddress,longbibliography]{revtex4-1}
\pdfoutput=1
\usepackage{url}
\usepackage{extarrows}
\usepackage{hyperref}
\usepackage{epsfig}
\usepackage{amsmath}
\usepackage{graphicx}
\usepackage{wrapfig}
\usepackage{dcolumn}
\usepackage{bm}
\usepackage{amssymb}
\usepackage{titlesec}
\usepackage{mathtools}
\usepackage{relsize}
\usepackage{cleveref}
\usepackage{calligra}
\usepackage{lipsum}
\usepackage{xcolor}
\usepackage[T1]{fontenc}
\usepackage[latin9]{inputenc} 
\usepackage{geometry}
\usepackage{natbib}
\usepackage{babel}
\usepackage{lineno}
\begin{document}

\title{Target-density formation in swarms with stochastic sensing and dynamics}
\author{Jason Hindes$^{1}$, George Stantchev$^{1}$, Klimka Szwaykowska Kasraie$^{2}$, and Ira B. Schwartz$^{1}$}
\affiliation{$^{1}$U.S. Naval Research Laboratory, Washington, DC 20375, USA}
\affiliation{$^{2}$Georgia Tech Research Institute, Atlanta, GA, 30318, USA}

\begin{abstract}
An important goal for swarming research is to create methods for predicting, controlling and designing swarms, which produce collective dynamics that solve a problem through emergent and stable pattern formation, without the need for constant intervention, and with a minimal number of parameters and controls. 
One such problem involves a swarm collectively producing a desired (target) density through local sensing, motion, and interactions in a domain. 
Here, we take a statistical physics perspective and develop and analyze a model wherein agents move in a stochastic walk over a networked domain, so as to reduce the error between the swarm density and the target, based on local, random, and uncertain measurements of the current density by the swarming agents.  
Using a combination of mean-field, small-fluctuation, and finite-number analysis, we are able to quantify how close and how fast a swarm comes to producing a target as a function of sensing uncertainty, stochastic collision rates, numbers of agents, and spatial variation of the target.    
%The approach has severl advantegs   
\end{abstract}
\maketitle

\section{\label{sec:I}INTRODUCTION}
%\begin{itemize}
    %\item Swarming review, culminating in designing-swarms to do-something. 
    %\item Important example: coverage and targeted pattern formation, survey 
    %\item Basic physical questions remain for truly autonomous swarms of simple, mobile agents. Emergent target formation in high-dimensional system from tuning a few external "knobs." Summary of results 
%\end{itemize}

%\textcolor{black}{Swarming in physics and biology}

Swarms consist of large numbers of self-propelled agents that interact to produce a wide variety of complex, chaotic, and coherent spatiotemporal behaviors\cite{VICSEK201271}. Typically, swarms are nonequilibrium systems in which agents consume energy in order to propel themselves in space and exhibit collective dynamics without central orchestration\cite{ActiveMatter}. 
Physical and biological examples have been observed across many space and time scales from colloidal swarms\cite{nano13101687,mi9020088} to colonies of bacteria \cite{Copeland2009BacterialSA,StatisticalPhysicsOfBacteriaSwarms}, large groups of insects\cite{Theraulaz2002SpatialPI,EquationOfStateForInsectSwarms,10.1371/journal.pcbi.1002642}, flocks of birds\cite{10.1371/journal.pcbi.1002894,Ballerini2007InteractionRA}, schools of fish\cite{10.1371/journal.pcbi.1002915,Calovi_2014}, and crowds of people\cite{PhysRevE.75.046109,ForceBasedPedestrianModels}. Much work has demonstrated how the collective dynamics of swarms can emerge through physically and biologically-inspired mechanisms and interactions\cite{PhysRevLett.75.1226,PhysRevE.63.017101,PhysRevLett.96.104302,4200853,Choi2017,SwarmShedding,PhysRevE.101.042202,Hindes_2024}. 

%Because of the robustness and scalability of such swarms, combined with the ever-expanding availability of mobile robotic platforms, swarm robotics is a field with great interest to the DoD. For instance, it is known that swarms of mobile robots can be designed and parameterized to fulfill defense missions, such as perimeter watch, landing site surveillance and protection, damage protection, 
%and intelligence, surveillance, and reconnaissance (ISR) missions to support the warfighter in maintaining
%situational awareness and closing the observe-orient-decide-act (OODA) loop. By tuning parameters for the
%entire swarm to create desired collective behaviors, the warfighter is freed from the need to individually
%control multiple assets, freeing attention for other tasks.

Because of the robustness and scalability of biological and physical swarms, and the continual advancement of mobile robotic platforms and capabilities\cite{1ff29b47f3ba40e6a5d74174084fc01a,9460560,Mezey2025}, there is great interest in designing robotic swarms to perform collective missions in defense and industry\cite{8722887,CooperativeLocalizationAndMapping,4542870,6733749,DroneSwarmStrategyForDetectionAndTracking,9432146,8206299,s22134773,PhysRevE.103.062602}, and even physics\cite{Liu2023}. In addition to target tracking and flocking\cite{4200853,doi:10.1049/ietcta:20050401,doi:10.2514/1.25629,Li2018,Mezey2025}, a canonical problem for mission-driven swarms pertains to optimal coverage over a domain\cite{1284411,7798696,ELAMVAZHUTHI2018356} and prescribed density formation\cite{5990615,EREN20179405}. Of interest to us is the latter, where a swarm evolves to produce a particular density profile in space. Most works on the subject build optimal motion controllers for swarming agents based on solutions to advection-diffusion equations\cite{EREN20179405,9483288,9525253,SINIGAGLIA2025112218} or optimized Markov chains\cite{4926163,6314729,CUI2024111832,7948777}, and prove convergence of a swarm to the desired density under a variety of conditions and assumptions. Particularly interesting are mean-field control methods with density estimation\cite{EREN20179405,9525253,9483288} and static optimal control\cite{SINIGAGLIA2025112218}, as well as discrete-state and time models showing self-repair\cite{6314729}. For success, however, most take a tightly engineered perspective, and work from limits where the swarming agents rely on complex computing capabilities {\it somewhere} -- either offline by a central orchestrator, or by agents themselves, which are designed with the capacity to perform intricate calculations, and/or sense and communicate their kinematic data to other agents in a swarm with high fidelity. Moreover, in many models the agents in a swarm do not directly interact at all, and so the effects of basic physical processes and stochasticity are unknown. 

One of the anticipated advantages of a true swarming system is the ability to solve a collective dynamics problem where the mission-driven behavior is emergent from the interactions of simple and limited agents, and can be changed and stabilized through a relatively small number of ``knobs" or parameters, and without constant external intervention  at the level of each agent. Currently, lacking is a basic physical approach to targeted density formation, providing first-principles, quantitative answers for such questions as: how close does a swarm's emergent density come to a targeted pattern and how fast can it produce such a pattern, given simple random and stochastic dynamics for the agents and awareness of the target, and how do these answers depend on the spatial complexity of the target, the sensing uncertainty and stochasticity, the numbers of agents, the rate of collisions among the agents, etc.? It is just such an approach that we build in this work, and thereby answer the questions posed. In particular, we formulate and analyze a model where agents move stochastically over a domain with the goal of reducing the error between a known target and the current swarm density. The latter is perceived by a given agent on the basis of random and uncertain local measurements. In so doing, we offer a starting point for future investigation and analysis from a physics perspective on the problem of autonomous targeted spatial-temporal density formation.

Our paper is organized as follows. In Sec.\ref{sec:SSD} we introduce the stochastic swarming model and its mean-field theory. In Sec.\ref{sec:SSC} we perform analysis of the dynamics in several important limits and determine cross-over regimes that separate different behaviors. In Sec.\ref{sec:D} we provide a discussion of the results and thoughts on generalizations.

\section{\label{sec:SSD}STOCHASTIC SWARM DYNAMICS}
We are interested in studying swarms that build target densities in space from stochastic and local dynamics with minimal control. In particular, \textcolor{black}{we focus on the spatial allocation of swarms with a fixed number of agents, rather than swarms whose sizes fluctuate in time, or swarms that target an absolute number of agents per unit volume.} For this goal, a useful starting point is to construct continuous-time Markov processes for the essential physics, which naturally incorporate stochasticity and locality\cite{Kampen1992,Masuda_Vestergaard_2023}. Consider a swarm of $N$ agents in which every agent can move in a domain partitioned into $M$ fixed subregions (or patches), over which a target density is to be constructed. In general, the patches and the connections among them can be viewed respectively as the nodes and edges of a graph specified by an adjacency matrix, $A$, where $A_{ij}\!=A_{ji}\!=\!1$ if patches $i$ and $j$ are connected, and $A_{ij}\!=\!A_{ji}\!=\!0$ otherwise. 
%\GMS{Would be nice to see a bit more formal definition of the dynamics, formulated as discrete-time mean-field stochastic process, similar in spirit, say to how a particle-based Markov Chain Monte Carlo algorithm would be described, sans the gory statistical details. Just a couple of sentences outlining the mathematical framework, before moving to the description of the specific iterative scheme. Also should emphasize that $t$ is discrete "time". I can take a stab at this tomorrow}
Furthermore, let us denote the swarm density at patch $i$ at time $t$, $y_{i}(t)\!=\!n_{i}(t)/N$, where $n_{i}(t)$ is the current number of agents located in patch $i\!\in\!\{1,2,...,M\}$. \textcolor{black}{Note that by `density' we refer to the fraction of a fixed-size swarm that occupies a given patch, where domain patches have constant and equal volumes.} The goal of the swarm is to collectively produce a certain target density profile, 
$\{\overline{y}_{1},\overline{y}_{2},...,
\overline{y}_{M}\}$. 
%represented as a partition $\{\overline{y}_{i}\}$ over patches $i\in\{1,2,...,M\}$. 
In this work, we assume that all agents know the full target profile, and they attempt to create it using local sensing and movement through the domain. 
%\GMS{I would suggest changing the notation for the patch densities in a way that distinguishes better the actual and observed density for an explicitly given agent. For example, the swarm density over patch $i$ at time $t$ could be denoted $y^{i}_t\!=\!n^{i}(t)/N$, while the density over patch $i$ observed by agent $s$ could become $y^{i}(s)$. That would avoid the use of $'$ to denote a specific agent, which can be confusing. Also when time is used as subscript (which is natural in many contexts), the target density over patch $i$ would naturally be written as $y^i_T$. [GS]}

In particular, in order to build the target density agents measure their local patch density and the patch density at a {\it randomly} selected neighboring patch: both with some measurement uncertainty. Based on their measurements, agents move to reduce the perceived error with the target. To that end, let us assume that every agent makes its joint density measurements with probability per unit time $\alpha$, independently of the rest of the swarm so that there is no assumed synchronization of measurements among agents. For a given agent $\Omega \in [1,N]$ let $i$ be the patch that it occupies at time $t$, and let a patch $j$ be selected uniformly at random from the neighbors of patch $i$ with probability $1/k_{i}$, where $k_{i}=\sum_{j}A_{ij}$. Let $\hat{y}_{i}$ and $\hat{y}_{j}$ denote agent $\Omega$'s density measurements for patches $i$ and $j$, respectively. 
Given $\hat{y}_{i}$ and $\hat{y}_{j}$, $\Omega$ attempts to bring the swarm closer to the target by moving to patch $j$, if $\overline{y}_{j}\!-\!\hat{y}_{j} \geq  \overline{y}_{i}\!-\!\hat{y}_{i}$, or staying in patch $i$ otherwise. Altogether, if we define the random variable $\hat{z}_{ij}\!=\!\overline{y}_{i}-\hat{y}_{i}-\overline{y}_{j}+\hat{y}_{j}$,\footnote{If an agent at patch $i$ moves to $j$, the change in the squared Euclidean distance to the target is $2\hat{z}_{ij}/N$ for large $N$, as perceived by the agent.}
with a \textcolor{black}{probability density Pr($\hat{z}_{ij}$ ) and cumulative distribution function $C(\hat{z}_{ij})\!=\!\int_{-\infty}^{\hat{z}_{ij}}Pr(\hat{z}_{ij}')d\hat{z}_{ij}'$}, then the rate for the continuous-time Markov process, describing agent $\Omega$'s discrete movement from $i$ to $j$, is $\alpha C(\hat{z}_{ij}=0)/k_{i}$. Hence, in terms of the swarm's density, we have the following stochastic reaction associated with $\Omega$'s movement from $i$ to $j$:
%\GMS{ which is used to determine the transition rate of a stochastic Markov process describing agent $\Xi$'s discrete trajectory as a sequence of moves from one patch to a neighboring one. In particular, there is an associated stochastic Markov reaction}
\begin{align}
\label{eq:move}
\!\!(y_{i},y_{j})\;\;\xlongrightarrow[]{\;\bigl(\alpha C(\hat{z}_{ij}=0)/k_{i}\bigr)\;}\;\;(y_{i}-\frac{1}{N},y_{j}+\frac{1}{N}).
\end{align}
%\GMS{in terms of density variation in patches $i$ and $j$ induced by the stochastic motion of $\Xi$}.
In general, the uncertainty in measurement is encapsulated in $C(\hat{z}_{ij})$, which is a function of the true and target densities at patches $i$ and $j$. This function depends on the physics of measurement, and in this work we assume that it has the same functional form for all agents. %\IBS{I also am not a fan of the superscript '. I would make the uncertain measurements notated with a carrot top, making it easier to read. Subscripting "t" as a time variable. assumes it is a parameter, and since there is no real clock here, we can consider the original notation for time suppressed; i.e., dropping the variable for convenience.}\GMS{Agreed, time can be suppressed here, so let's suppress it}
 
\begin{figure*}[t]
\center{\includegraphics[scale=0.165]{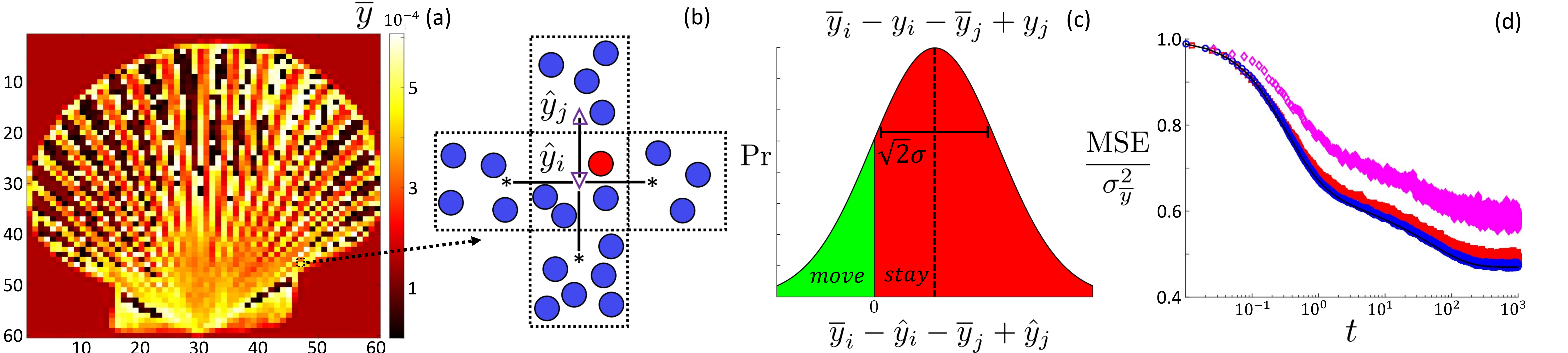}}
\caption{Swarm evolving toward a target density through local sensing and dynamics. (a) The target known to the agents. (b) An example patch and its four neighbors. A randomly selected agent shown in red, makes two uncertain density measurements-- one at its local patch $i$, and another at a randomly selected neighboring patch $j$. (c) \textcolor{black}{The probability density (Pr) for how far-off the densities of the patches are from their target, $\hat{z}_{ij}=\overline{y}_{i}\!-\!\hat{y}_{i}\!-\!\overline{y}_{j}\!+\!\hat{y}_{j}$, as estimated by the red agent}. In the example, the probability density is \textcolor{black}{Gaussian with mean $\overline{y}_{i}\!-\!y_{i}\!-\!\overline{y}_{j}\!+\!y_{j}$ and variance $2\sigma^{2}$}. If the difference between the neighboring patch's estimated density and the target is greater than the difference for its current patch, the agent moves to the neighboring patch. (d) swarm density mean-squared error (divided by the target variance) versus time for: $N=$ $10^4$ (magenta), $10^5$ (red), and $10^6$ (blue). The solution of Eq.(\ref{eq:MFpatternFormation}) is plotted with a black line.  \textcolor{black}{Pr($\hat{z}_{ij}$) is the same as (c)}. Other parameters are: $M\!=\!60^{2}$, $\sigma M\!=\!0.5$, and $\beta\!=\!1$.}
\label{fig1}
\end{figure*}

An illustration of the stochastic dynamics through which a swarm attempts to produce a target density is shown in Fig.\ref{fig1}. Panel (a) plots a heatmap of the target density, which is built over a periodic square lattice with $M=60^{2}$ and $k_{i}\!=\!4\;\forall i$. Panel (b) shows an example patch where an agent in red makes measurements of its local density $\hat{y}_{i}$ and the density at a single, randomly selected neighboring patch $\hat{y}_{j}$. Depending on the outcome of these measurements, the agent moves to the neighboring patch or stays at the current patch. Panel (c) plots an example, Gaussian probability distribution for the measurement outcomes, which has a mean $\overline{y}_{j}-y_{j}-\overline{y}_{i}+y_{i}$ and standard deviation $\sigma$.    

%Furthermore, each agent can sense the swarm density at its current location (i.e, if an agent is located at patch $i$, it can measure $y_{i}(t)$ at time $t$), and at a single ``neighboring" patch: both with some measurement uncertainty. 
%In order to build the target density, agents randomly select a neighboring patch and perform their density measurements. 
%By assumption, the process occurs independently with a probability per unit time $\alpha$ for each agent. Let us denote an example agent's current density estimate, $y_{i}^{'}$, for patch $i$, and $y_{j}^{'}$ for a neighboring patch $j$. Based on its uncertain measurements, the agent attempts to bring the swarm closer to the target by moving to patch $j$, if $y_{j}^{T}\!-\!y_{j}^{'} >  y_{i}^{T}\!-\!y_{i}^{'}$. 

In addition, to the sensing-based dynamics, one can consider other basic physical ingredients such as collision between agents in the same patch. For example, as agents enter and leave a patch, they may jostle nearby agents, resulting in random motion in the swarm. Let us assume that each agent occupies a characteristic volume, $w$, and if two agents overlap within a patch, one of them tends to be ejected at a rate $\gamma$ to a randomly chosen neighboring patch, from recoil. In addition, if we assume that agents within a patch are roughly uniformly distributed in space, the overlap probability between a given agent and any other at patch $i$, is approximately $(w/V)Ny_{i}$, where $V$ is a characteristic patch volume. Therefore, similar to the sensing-based motion, an agent $\Omega$ at patch $i$ will undergo a stochastic Markov collision-ejection reaction to patch $j$ at a rate $M\beta y_i /k_{i}$, where $\beta\!\equiv\!\gamma wN\!/(MV)$ is a 
rate constant
\footnote{Note that in order for $\beta$ to remain finite as $N\!\rightarrow\!\infty$ for fixed $M$ and $w$, $V$ should scale linearly with $N$. In addition, if ejection occurs only when collisions are near a patch boundary, one can reduce $\beta$ by the appropriate fraction of the patch volume that is close enough to a boundary to result in ejection}. 
Written in terms of the swarm density, the reaction is 
%agent $\Xi$ at patch $i$ undergoes a stochastic Markov collision-ejection reaction \GMS{with transition rate $M\beta y_i$, which gives rise to stochastic density variation in patches $i$ and $j$}
\begin{align}
\label{eq:repul}
(y_{i},y_{j})\;\;\xlongrightarrow[]{\;\bigl(M\beta y_{i}/k_{i}\bigr)\;}\;\;(y_{i}-\frac{1}{N},y_{j}+\frac{1}{N}). 
\end{align}

In summary, every agent at patch $i$ undergoes measurement and collision reactions to a single neighboring patch $j$ according to Eqs.(\ref{eq:move}-\ref{eq:repul}), (with a total of $2k_{i}$ reactions for each agent at patch $i$). To simulate the dynamics of the whole swarm, one can generate reaction times stochastically using Gillespie's algorithm\cite{Masuda_Vestergaard_2023} for all agents at all patches, and for example, select the reaction that occurs first before repeating. Given the dynamics of this set of physical reactions, one would like to know how the swarm evolves toward the target, how long it takes to reach a steady-state, what the swarm density error is compared to the target,  how does the swarm density depend on the number of agents, etc. 

To make progress, we approach the dynamics in the manner of statistical physics, and first consider its mean-field behavior, valid in the limit of large numbers of agents.
The mean-field dynamical system can be derived from the Markov-processes specified in Eqs.(\ref{eq:move}-\ref{eq:repul}), by setting the time derivatives of the swarm densities equal to the sum over the rates multiplied by the increments for all possible reactions that change a given density\cite{10.1063/1.467139}. For instance,  the contribution to $dy_{i}/dt$ from agents at $i$ leaving for patch $j$, because of measurement, is the product of three terms: the rate at which a single agent leaves $(\alpha C(\hat{z}_{ij}=0)/k_{i})$, the change to $y_{i}$ when an agent leaves $(-1/N)$, and the number of agents at patch $i$ $(n_{i})$.  
%rate of a single agent at $i$ leaving, multiplied by the change to $y_{i}$, multiplied by the number of agents at patch $i$: $\alpha C(\hat{z}_{ij}=0)/k_{i}\cdot(-1/N)\cdot n_{i}$, respectively.
%\GMS{Perhaps a better way to say this would be: the contribution to $\frac{dy_i}{dt}$ from agents at $i$ entering/leaving for/coming from patch $j$, because of measurement, is the product of the probability of a single agent at $i$ leaving, the signed change to $y_{i}$, and the number of agents at patch $i$, namely: $\alpha C(z_{ij}^{'}=0)/k_{i}, (\pm 1/N)$, and $n_{i}$, respectively}
By adding similar terms for agents entering patch $i$ from patch $j$, summing over all neighbors of $i$, and repeating for movement resulting from collision/repulsion, we find
\begin{align}
%\dot{y}_{i}
\dfrac{d{y}_{i}}{\!\!dt}=&\;\alpha\sum_{j}A_{ij}\Big[\frac{y_{j}}{k_{j}}C(\hat{z}_{ji}=0)-\frac{y_{i}}{k_{i}}C(\hat{z}_{ij}=0)\Big] \nonumber \\
&+M \beta\sum_{j}A_{ij}\Big[\frac{y_{j}^{2}}{k_{j}}-\frac{y_{i}^{2}}{k_{i}}\Big]. 
\label{eq:MFpatternFormation}
\end{align}
We can analyze the system Eqs.(\ref{eq:MFpatternFormation}) in order to understand in detail how a large swarm evolves toward a target density.

Figure \ref{fig1} (d) shows an example comparison between the dynamics of Eq.(\ref{eq:MFpatternFormation}), plotted with a black line, and stochastic simulations of swarms with $10^4$ (magenta), $10^5$ (red), and $10^6$ (blue) agents. Plotted is the swarm's mean-squared error $\text{MSE}(t)\!=\!\sum_{i}\left(\overline{y}_{i}\!-\!y_{i}(t)\right)^{2}\!\!/M$, normalized by the spatial variance of the target $\sigma^{2}_{\overline{y}}\!=\!\sum_{i}\left(\overline{y}_{i}\!-\!1/M\right)^{2}\!\!/M$. As we expect, the stochastic dynamics approaches the mean-field as $N\!\rightarrow\!\infty$.

\section{\label{sec:SSC}STABILITY, SCALING, AND CROSSOVER}
The dynamics of Eqs.(\ref{eq:MFpatternFormation}) are difficult to study in full generality. 
%, apart from numerical simulation. 
However, basic insight can be gained %into the behavior 
by first focusing on a simple model for the measurement uncertainty and assume that patch measurements are independent Gaussian processes, whose means are given by the true density and whose standard deviations are constant, $\hat{y}_{i}\sim G(y_{i},\sigma/\sqrt{2})$. Consequently, $C(\hat{z}_{ij}\!=\!0)\!=\!(1+\text{erf}\{[\overline{y}_{j}\!-\!y_{j}\!-\!\overline{y}_{i}\!+\!y_{i}]/[\sqrt{2}\sigma]\})/2$, where `erf' denotes the error function. Of course, other %measurement 
models can be considered; see Sec.\ref{sec:D} for discussion. Also, note that for this model, if agents make multiple ($m$) measurements, the error is simply rescaled as the standard measurement error, $\sigma\rightarrow\sigma/\sqrt{m}$. In addition, let us assume that the graph is approximately $k$-regular, $k_{i}\!\cong\!k\;\forall i $; two relevant graph classes are periodic lattices and random networks with homogeneous degree\cite{newman2018networks}. The general graph case is treated in App.\ref{sec:A} with similar results. 

Given these assumptions, it is useful to first study the limit where the target density has relatively small spatial fluctuation. \textcolor{black}{If we define  
$\epsilon\!\equiv\!M\sigma_{\overline{y}}$, we can define a target fluctuation, $\;\overline{\!\!f}_{i}\equiv (M\overline{y}_{i}-1)/\epsilon$, which measures deviation from a uniform density profile.} If $\epsilon\ll1$, then we expect the swarm density to take the form of a power series in $\epsilon$ with \textcolor{black}{coefficients $f_{i,n}\equiv\frac{1}{n!}\frac{\partial^{n} y_{i}}{\partial\epsilon^{n}}|_{\epsilon=0}$}, or
%Finally, for analysis, we assume that the target density has relatively small spatial fluctuation such that $\overline{y}_{i}\!=\!(1+\epsilon\;\overline{\!\!f}_{i})/M$, where $\epsilon\!\equiv\!M\sigma_{\overline{y}}$. Note that we can write the target in this form in general, but when $\epsilon$ is small, one can perform an expansion of the swarm density,
\begin{equation}
\label{eq:expansion}
y_{i}(t)=(1+\epsilon f_{i,1}(t)+\epsilon^{2} f_{i,2}(t)\;+...)/M.   
\end{equation}

\subsection{\label{sec:SF}Small fluctuations}
To understand the small fluctuation (SF) dynamics, we substitute Eq.(\ref{eq:expansion}) into Eqs.(\ref{eq:MFpatternFormation}), and collect powers in $\epsilon$. Defining the vectors \textcolor{black}{$\overline{\!\!f}\equiv (\;\overline{\!\!f}_{\!1},\;\overline{\!\!f}_{\!2},...,\;\overline{\!\!f}_{\!M})$}
and \textcolor{black}{$f_n\equiv(f_{1,n},f_{2,n},...,f_{M,n})$}, 
at $\mathcal{O}(\epsilon)$ the result is 
%At $\mathcal{O}(\epsilon)$ the result is  
\begin{align}
\frac{df_1}{dt}= \Big(\frac{\alpha}{2}+2\beta+\frac{2\alpha}{\sqrt{2\pi}M\sigma}\Big) Lf_1 -\frac{2\alpha}{\sqrt{2\pi}M\sigma} L\;\overline{\!\!f},
%\dot{f}_{i,1}=&\Big(\frac{\alpha}{2}+2\beta+\frac{2\alpha}{\sqrt{2\pi}M\sigma}\Big)\sum_{j}A_{ij}[f_{j,1}-f_{i,1}] \nonumber\\
%&-\frac{2\alpha}{\sqrt{2\pi}M\sigma}\sum_{j}A_{ij}[f_{j}^{T}-f_{i}^{T}]. 
\label{eq:FirstOrderKreg}
\end{align}
where $L$ is the graph Laplacian, $L=A-D$, and $D$ is the diagonal degree matrix, $D_{ij}=k_{i}\delta_{ij}$ \cite{mieghem_2010}. 
%\GMS{
%Is the component-wise notation necessary here? How about writing Equ.~\ref{eq:FirstOrderKreg} as a vector equation like this: 
%\[
%\frac{df_1}{dt}=\xi Lf_1 + \eta Lf^T
%\]
%where $f_1$ and $f^T$ are now vectors with components $f_{i,1}$ and $f_i^T$ respectively, $i=1,\dots,M$, $L$ is the graph Laplacian, $\xi = \frac{\alpha}{2}+2\beta+\frac{2\alpha}{\sqrt{2\pi}M\sigma}$, and $\eta = -\frac{2\alpha}{\sqrt{2\pi}M\sigma}$
%}
%\GMS{By the same token the eigenvector of the Laplacian associated with eigenvalue $\mu_l$ can be denoted by $v_l$ and defined by $Lv_l = \mu v_l$, thus avoiding the component-wise notation. Similarly the mode decomposition of $f_1$ could be written in terms of scalar products of $v_l$ and $f_1$ using for instance the $<,>$ notation. This is not critical, I'm just expressing my personal preference for vector/global notation whenever component-wise/local notation is not needed -- makes for more concise, conceptually cleaner notation
%}
Equation (\ref{eq:FirstOrderKreg}) can be easily decomposed into the modes of $L$. Let us denote the eigenvalues $\{\mu_{l}\}$ and right eigenvectors $\{v_{l}\}$ of the Laplacian, $\mu_{l}v_{l}=Lv_{l}$.  The density projections onto the modes are $c_{l}=v_{l}^{\intercal}f_{1}$ and  
$\overline{\!c}_{l}=v_{l}^{\intercal}\;\overline{\!\!f}$, respectively, where $\intercal$ is the transpose operation. In terms of the projection,  solutions to Eq.(\ref{eq:FirstOrderKreg}) take the simple form:
\begin{subequations}
\begin{align}
\label{eq:LinearSolution1}
 c_{l}(t)&=&\big(c_{l}(t=0)-B_{l}\big)e^{\lambda_{l}t}+B_{l},\\
 \label{eq:LinearSolution2}
 \lambda_{l}&=&\frac{\mu_{l}}{k}\Big(\frac{\alpha}{2}+2\beta+\frac{2\alpha}{\sqrt{2\pi}M\sigma} \Big),\\
 \label{eq:LinearSolution3}
 B_{l}&=& \dfrac{\dfrac{2\alpha}{\sqrt{2\pi}M\sigma}}{\dfrac{\alpha}{2}+2\beta+\dfrac{2\alpha}{\sqrt{2\pi}M\sigma}}\;\;\overline{\!c}_{l}\;. 
\end{align}
\end{subequations}

Several important insights follow from Eqs.(\ref{eq:LinearSolution1}-\ref{eq:LinearSolution3}). We start with dynamics. First, the approach to the steady-state is monotonic with timescales inversely proportional to the graph Laplacian eigenvalues (and proportional to $k$). In general, for a connected graph we can order the eigenvalues such that $0\!\geq\! \mu_{M-1}\!\geq\! \mu_{M-2}\!\geq\! ...\!\geq\! \mu_{1}$ \cite{mieghem_2010}. We point out that the homogeneous mode, $v_{i,M}\!=\!1/\sqrt{M}$ with $\mu_{M}\!=\!0$, plays no role in the dynamics, since $\sum_{i}y_{i}(t)\!=\!1$ implies $c_{M}(t)\!=\!0$. As a consequence, the steady-state density produced by the swarm is unique within the SF approximation, since the initial-condition dependence of Eq.(\ref{eq:LinearSolution1}) decays away. Furthermore, the characteristic rate over which the steady-state is reached is determined by the %\GMS{Fiedler} 
Fiedler eigenvalue of the graph $|\mu_{M-1}|$, which is {\it independent of the target properties}. Hence, we expect swarms to take approximately the same amount of time to produce different target patterns, given the same physical parameters. In terms of such parameters, increasing the measurement precision (decreasing $\sigma$) increases the speed at which a swarm changes its density, as does increasing the rate of collision, $\beta$. 

\begin{figure}[h]
\center{\includegraphics[scale=0.154]{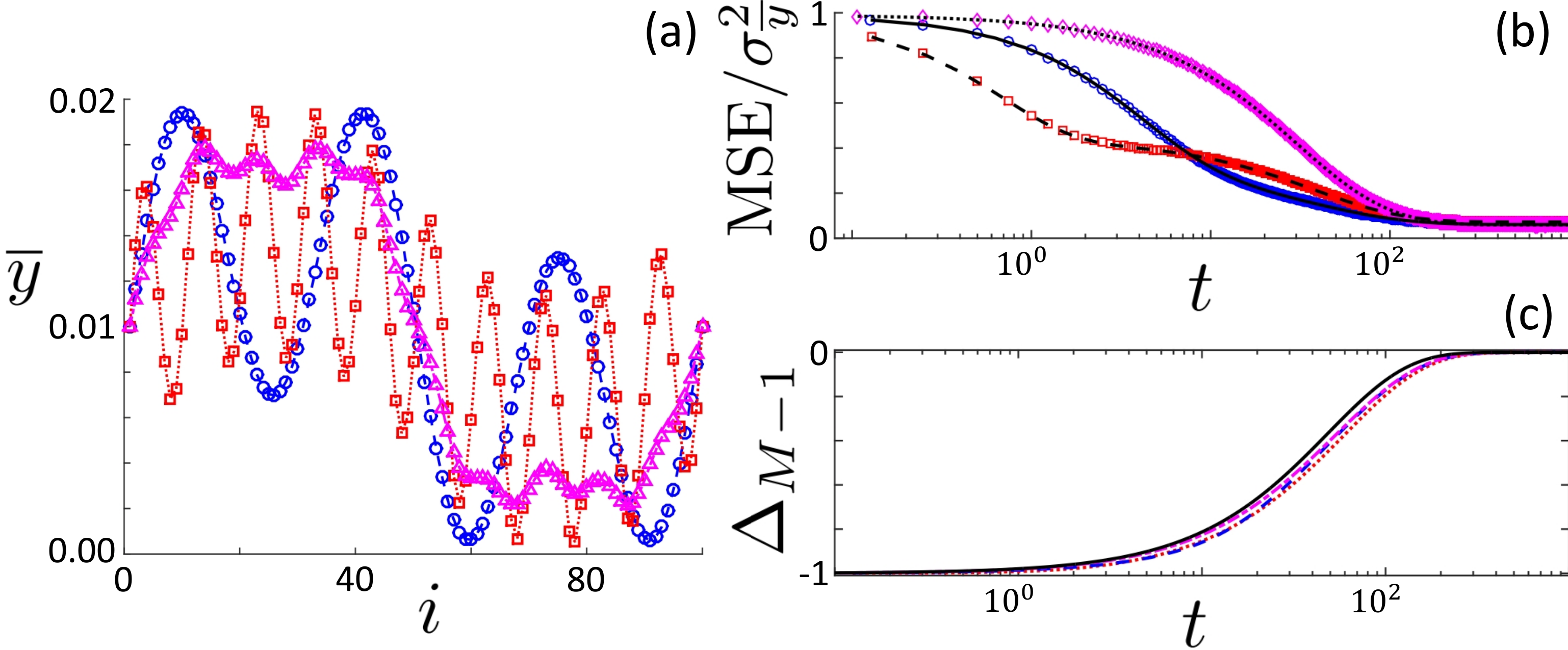}}
\caption{Dynamics of density formation. (a) Three target densities, each plotted with a different color. All panels follow the same color convention. (b) Stochastic simulations with $N\!=\!10^{4}$. %are plotted with circles, squares, and diamonds, respectively. 
Lines correspond to solutions of Eqs.(\ref{eq:MFpatternFormation}). The mean-sqarred error is plotted versus time. (c) Projection of the dynamics onto the Fiedler mode. Solutions of Eqs.(\ref{eq:MFpatternFormation}) for each target compared to Eqs.(\ref{eq:LinearSolution1}-\ref{eq:LinearSolution3}) for $c_{M-1}(t)$. Other parameters are: $M\!=\!100$, $\sigma M\!=\!0.1$, and $\beta\!=\!1$.}
\label{fig2}
\end{figure}

To verify our conclusions, we compare the dynamics of a swarm attempting to produce three different target densities over a periodic 1-d lattice with $k\!=\!2$ in Fig.\ref{fig2}. Each target is a superposition of three sine waves (with spatial periods $M$, $M/3$, and $M/10$), each with different sets of random amplitudes. The targets are plotted in Fig.\ref{fig2} (a). Note that $\epsilon\sim1$ in each case. In addition, the initial swarm density is $y_{i}(t\!=\!0)\!=\!1/M\;\forall i$. In panel (b), we show the mean-squared error ($\text{MSE}$) of the swarm density compared to the target, normalized by the spatial variance of the target $\sigma^{2}_{\overline{y}}$. 
Panel (b) demonstrates that different trajectories
emerge for each target pattern. However, the time it takes for the swarm to reach steady-state and the steady-state MSE normalized by the target variance are nearly identical for all targets. As pointed out, this is a prediction of the SF theory. In fact, we can compare the dynamics along the Fiedler mode of the graph in each case to the SF approximation. Figure \ref{fig2} panel (c) plots the relative distance to steady-state, $\Delta_{M-1}\!=-1+v_{M-1}^{\intercal}y/v_{M-1}^{\intercal}y (t\!\rightarrow\!\infty)$
%$\Delta_{M-1}(t)\!=\!\sum_{i}\!v_{i,M-1}[y_{i}(t)\!-\!y_{i}(t\!\rightarrow\!\infty)]/\sum_{i'}v_{i',M-1}y_{i'}(t\!\rightarrow\!\infty)$ 
for each example.
%\GMS{Nice switching to $\intercal$ for the transpose, but I still think you mean $v_{M-1}^{\intercal}y/v_{M-1}^{\intercal}y^T$ instead of $v_{M-1}^{\intercal}y/v_{M-1}^{\intercal}y$, no?}
%\GMS{Did you mean $v_{M-1}^{\intercal}y/v_{M-1}^{\intercal}y^T$ ? Also, transpose is more commonly typeset with $\intercal$ (\${\textbackslash}intercal\$) rather than 
%$\intercal$, and in this case if for no other reason, 
%$\intecal$ looks more distinct from superscript $T$ used for target densities and such}
Despite the significant spatial variation in the targets, 
each example closely follows the SF trajectory Eqs.(\ref{eq:LinearSolution1}-\ref{eq:LinearSolution3}) for $c_{M-1}(t)$. 
%The results show very good agreement despite the large values of $\epsilon$.   

Next, we address how close a swarm comes to building a target for a given set of physical parameters. In particular, we are interested in the MSE as $t\!\rightarrow\!\infty$. Again, using the SF approximation Eqs.(\ref{eq:LinearSolution1}-\ref{eq:LinearSolution3}), and summing over the modes, we find
\begin{align}
\frac{\text{MSE}(t\!\rightarrow\!\infty)}{\sigma^{2}_{\overline{y}}}=\left(\dfrac{1+\dfrac{4 \beta}{\alpha}}{1+\dfrac{4 \beta}{\alpha}+\dfrac{4}{\sqrt{2\pi}M\sigma}}\right)^{2}\!\!\!.
\label{eq:SFsteady}
\end{align}
Equation (\ref{eq:SFsteady}) is intriguing for several reasons. First, the MSE is predicted to be proportional to the target variance. Namely, doubling the variance while keeping physical parameters constant, doubles the error. On the other hand, the normalized MSE, expressed in units of the target variance, is independent of target properties, and depends only on parameters. In terms of the measurement error $\sigma$, as $\sigma M\rightarrow \infty$ the normalized error goes to $1$, as we expect. On the other hand, as $\sigma M\rightarrow 0$, the normalized error goes to zero. Hence, for targets with small spatial variation, swarms will tend to reach the target without discrepancy in the limit of zero measurement uncertainty. Note that this is the case, even for finite repulsion. Within the SF approximation, increasing $\beta$ results in larger error, but does not change the limiting behavior: repulsion does not prevent reaching the target as measurement uncertainty is reduced. We return to this prediction in Sec.\ref{sec:repulsion} and see that it is violated for large repulsion and large spatial variation of the target.

\textcolor{black}{Up to this point, we have assumed that swarming agents are given an \textit{a priori} known target density and that their density measurements are uncorrelated. However, if the target density must also be estimated from external observations, as in biological mimicry\cite{10.1371/journal.pone.0256025}, then target uncertainty should be included as well. 
%a scenario where the target density may not be known exactly occurs when it must be estimated from external observations, as in biological mimicry. 
%In that case patterns perceived from the environment are reproduced by a collection of specialized epidermal cells that coordinate the local color of their associated chromatophores. 
Here, we point out that our analysis can easily accommodate both target-density uncertainty and spatial correlations in measurements when these effects are Gaussian distributed. 
%, if the former is also Gaussian distributed. 
The main feature that allows for this is the fact that 
%in this case 
every agent %would still 
makes a decision to move based on a linear combination of four quantities.
In particular, for an agent located at patch $i$ considering a move to patch $j$, the relevant variable is $\hat{z}_{ij}\!=\!\hat{\bar{y}}_{i}-\hat{y}_{i}-\hat{\bar{y}}_{j}+\hat{y}_{j}$, where $\hat{\bar{y}}_{i}$ and $\hat{\bar{y}}_{j}$ denote the estimated target densities for patches $i$ and $j$, respectively. Note that the only update to $\hat{z}_{ij}$ from Sec.\ref{sec:SSD} is that the target densities are now estimates and therefore carry $\!$`{\footnotesize$\wedge$}'$\!$ notation as well. Let us assume that the target knowledge and swarm-density measurements are independent from each other, but separately, multivariate Gaussian distributed between patches-- allowing for correlation between two patches. If the means are given by the true target and swarm densities, and the covariances are $\text{cov}(\hat{\bar{y}}_{i}\;,\hat{\bar{y}}_{j})\!=p_{(t)}\sigma_{(t)}^{2}$ and 
$\text{cov}(\hat{y}_{i}\;,\hat{y}_{j})\!=p_{(m)}\sigma_{(m)}^{2}$, respectively, where $p_{(t)}$ and $p_{(m)}$ are the Pearson correlation coefficients for the target and swarm-density measurements with variances $\sigma_{(t)}^{2}$ and $\sigma_{(m)}^{2}$, then $\hat{z}_{ij}$ is a Gaussian random variable with mean $\left<\hat{z}_{ij}\right>\!=\!\bar{y}_{i}-y_{i}-\bar{y}_{j}+y_{j}$ and variance $\text{var}(\hat{z}_{ij})\!=\!2\sigma_{(t)}^{2}(1-p_{(t)})+2\sigma_{(m)}^{2}(1-p_{(m)})$. Hence, to incorporate both target uncertainty and spatial correlation in measurements, we need only redefine $2\sigma^{2}\rightarrow 2\sigma_{(t)}^{2}(1-p_{(t)})+2\sigma_{(m)}^{2}(1-p_{(m)})$.}    

\subsection{\label{sec:crossover}Finite-$N$ crossover}
\begin{figure}[h]
\center{\includegraphics[scale=0.158]{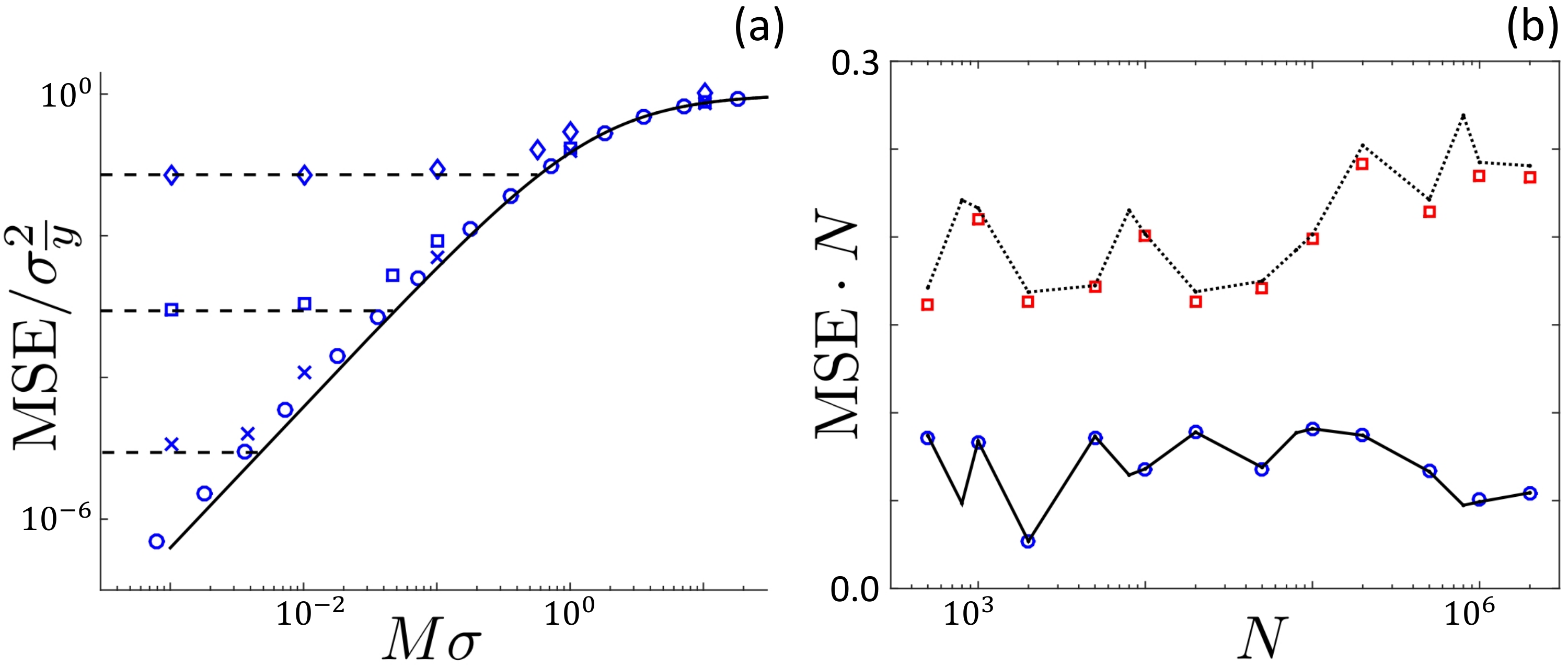}}
\caption{Finite-$N$ steady-state error. (a) $\text{MSE}$ versus $\sigma$ for: $N\!=\!10^4,$ (diamonds), $N\!=\!10^5,$ (squares), $N\!=\!10^6,$ (x's), and mean-field (circles). Equation (\ref{eq:SFsteady}) is plotted with a black line, while Eq.(\ref{eq:FiniteNSimple}) predictions are plotted with dashed lines. The target density is shown in Fig.\ref{fig1} (a). (b) $\text{MSE}$ versus $N$ for two disconnected target densities with $\sigma\!=\!0$: blue circles ($S\!=\!9$) and red squares ($S\!=\!24$). Equation (\ref{eq:FiniteN}) predictions are plotted with solid and dashed lines, respectively. $\beta\!=\!0$ for both panels.}
\label{fig3}
\end{figure}

Here, it is reasonable to wonder how swarms with finite $N$ behave as $\sigma$ is varied, and if the SF behavior holds. To make the picture clearer, let us set $\beta\!=\!0$. We return to $\beta\!\neq\!0$ in Sec.\ref{sec:repulsion}. Figure \ref{fig3} (a) shows several MSE series at steady-state versus $\sigma$: $N\!=\!10^4,$ (diamonds), $N\!=\!10^5,$ (squares), $N\!=\!10^6,$ (x's), and mean-field (circles). The target density and graph are the same as Fig.\ref{fig1} (a). The general pattern is the following: For each value of $N$ and large values of $M\sigma$, the behavior closely tracks the SF theory, which is plotted with a black line. However, for small values of $M\sigma$, the finite-$N$ systems cross-over and saturate to limiting values of error. The larger the value of $N$, the smaller the value of $M\sigma$ at which the crossover occurs. Note that the mean-field solution of Eqs.(\ref{eq:MFpatternFormation}) continues to track the SF theory closely.

Our next goal is to understand the crossover analytically. Because discrepancies between finite-$N$ systems and the mean-field behavior occur as $M\sigma\!\rightarrow\!0$, we let $\sigma\!=\!0$. \textcolor{black}{First, it is important to note that there is a fundamental quantization error with respect to the target simply by rounding each patch to the nearest integer, $n_{i}\!=\!R_{i}(N\bar{y}_{i})$, where $R_{i}$ is, for example, a floor or ceiling function with $N\!=\!\sum_{i}n_{i}$. Note that no matter what the rounding pattern, quantization produces a mean-squared error for the swarm density, $\sum_{i}(\bar{y}_{i}-y_{i})^{2}\!/M=\sum_{i}(N\bar{y}_{i}-R_{i}(N\bar{y}_{i}))^{2}\!/MN^{2}\leq 1/N^{2}$. In effect, the quantization error sets a limit on the MSE for the swarm dynamics, which should be approached as the measurement uncertainty goes to zero, $\text{MSE}\!\lesssim\!\mathcal{O}(1/N^{2})$.} 

We can gain more precise analytical insight by calculating the MSE in the case of special target densities, i.e., simple patterns that are {\it {disconnected}}.   
In particular, let us suppose that the target density has only $S\!\ll\!M$ patches for which $\overline{y}_{s}\!>\!0$, while the others are zero. Moreover, the patches with non-zero target density are not directly connected to each other. For convenience, we reorder the indices of the patches so that the `first' $s\!=\!1,2,..,S$ patches correspond to those with non-zero target. For disconnected targets with $S\!\ll\!M$, the swarm dynamics without measurement error is comparatively simple and can be analyzed phenomenologically. Starting from a uniform distribution, density flows into the non-zero target patches monotonically and at the same rate until $n_{s}\!\cong\!\text{floor}(N\overline{y}_{s})$, where the \text{`floor'} function rounds its argument down to the nearest integer. Note that if the patches with non-zero target are connected, then density flow could be in or out of a patch. When the rounding condition is satisfied, the remaining residual agents randomly walk over the patches with zero target, since their placement has no effect on the error. 

Thus, in terms of the $\text{MSE}$, there are two contributions: fixed error from patches with non-zero target, and stochastic error from the random walk of residual agents over patches with zero target density. Summing over the former and averaging the contributions of the latter gives 
\begin{align}
%MSE \cdot MN^{2}=&\sum_{s}\!\big(Ny_{s}^{T}-\text{floor}(Ny_{s}^{T})\big)^{2} \nonumber \\ 
(\text{MSE})(MN^{2})=&\sum_{s}\!\big(N\overline{y}_{s}-\text{floor}(N\overline{y}_{s})\big)^{2} \nonumber \\ 
&+N_{r}\Big(1+\dfrac{N_{r}-1}{M-S}\Big), 
\label{eq:FiniteN}
\end{align}
where $N_{r}\!=\!N-\sum_{s}\text{floor}(N\overline{y}_{s})$ is the number of residual agents. %\GMS{I would suggest using MSE in upright text consistently throughout the manuscript to denote Mean Squared Error, as it is technically an operator (though we are suppressing the arguments for convenience). That way in formulas like the one above it would be interpreted as a composite symbol, distinct from say $M$ and $N$ which are parameter names}

To test the finite-$N$ result, we plot the MSE as a function of $N$ for two examples with disconnected targets that are built over a periodic 1d lattice with $M\!=\!100$ and $k\!=\!2$. The results are shown in Fig.\ref{fig3} (b). Both targets have $\{\overline{y}_{s}\}$ that are randomly generated from independent uniform distributions over $[0,2/S]$,
$\overline{y}_{s}\sim U(0,2/S)$, but with different $S$
\footnote{Given a random sample, the target densities are renormalized, $\overline{y}_{s}\rightarrow \overline{y}_{s}/\sum_{i} \overline{y}_{i}$, so that $\sum_{s}\overline{y}_{s}\!=\!1$}: 
blue circles ($S\!=\!9$) and red squares ($S\!=\!24$). We can see that the error fluctuates with $N$ in a seemingly random way. Nevertheless, Eq.(\ref{eq:FiniteN}) captures the behavior, particularly for smaller $S$. 

Alternatively, when averaging over disconnected target patterns, 
a simpler analytical structure appears. For instance, 
given the i.i.d uniform model used 
for $\{\overline{y}_{s}\}$ in Fig.\ref{fig3} (b), the expectation value of Eq.(\ref{eq:FiniteN}) becomes
\begin{align}
\left<\text{MSE}\right>= \dfrac{5S}{6M}\dfrac{1}{N^{2}}. 
\label{eq:FiniteNSimple}
\end{align}
Namely, Eq. (\ref{eq:FiniteNSimple}) gives the simple result that in the limit of zero sensing error and repulsion, the expectation value of the $\text{MSE}$ is determined by the fraction of patches that have non-zero target density, over the total number of agents squared.

Now, we are in a position to determine the crossover point for Fig.\ref{fig3}(a). Specifically, to find the measurement uncertainty that separates mean-field behavior from finite-$N$ effects, we set Eq.(\ref{eq:SFsteady}) equal to Eq.(\ref{eq:FiniteNSimple}), assume $\sigma\!\ll\!1$, and ignore order-one constants. The result is 
\begin{align}
\sigma_{cr}\cong\dfrac{S^{1/2}}{M^{3/2}N\sigma_{\overline{y}}}.
\label{eq:CrossOver}
\end{align}
For example, the predicted crossovers from Eq.(\ref{eq:CrossOver}) for the three finite-$N$ series in Fig.\ref{fig3}(a) are represented by points of intersection between the black and dashed curves.

To summarize: for large $N$ swarms, the density error with respect to the target is generally described by mean-field theory and SF scaling for measurement uncertainties satisfying $\sigma\!\gtrsim\!\sigma_{cr}$.
However, when $\sigma\!\lesssim\!\sigma_{cr}$ the error saturates to a limiting value. Hence, $\sigma_{cr}$ defines a practical lower-bound for the measurement dynamics, in that, for a fixed size $N$, smaller uncertainty will not produce swarm densities with less error relative to the target.

%\textcolor{black}{Check $S$ scaling}
%\IBS{Can we plot the location of the crossover $\sigma_{cr}$? Also, a bit more on its significance, since it seems almost parenthetical the way the section ends.}

\subsection{\label{sec:repulsion}Large repulsion}
So far we have seen that the SF theory captures the general quantitative scaling for swarm target-density formation under parameter variation. In particular, as the agent measurement error is reduced, a swarm tends to approach a target monotonically until a crossover $\text{MSE}\sim\mathcal{O}(N^{-2})$ is reached. However, if repulsion is too strong, the trend of reducing error by reducing $\sigma$ can be violated. To see this, note that when $\beta\!\gg\!\alpha$, the swarm dynamics is dominated by repulsion (even when $\sigma\!\rightarrow\!0$), which tends to produce a steady-state of Eqs.(\ref{eq:MFpatternFormation}), $y_{i}(t\!\rightarrow\!\infty)\!=\!\sqrt{k_{i}}/\sum_{j}\!\sqrt{k_{j}}$; for $k$-regular networks the result is $y_{i}\!\rightarrow\!1/M$ and $\text{MSE}/\sigma_{T}^{2}\!\rightarrow\!1$.     

\begin{figure}[h]
\center{\includegraphics[scale=0.182]{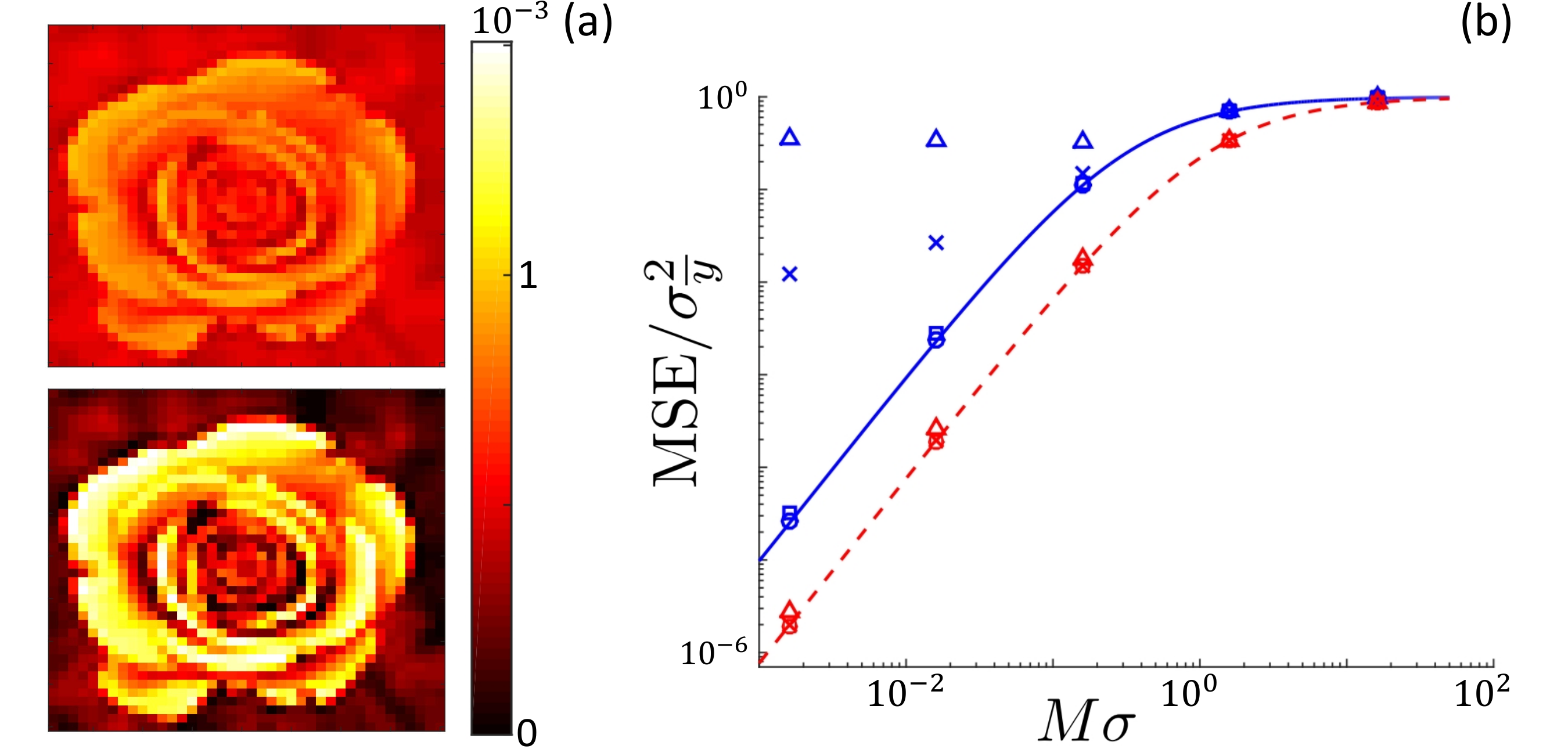}}
\caption{Dependence of the error on repulsion and spatial variation of the target. (a) Two example targets with small (top, $p\!=\!0.1$) and large (bottom, $p\!=\!1.0$) spatial variation. (b) $\text{MSE}$ versus $\sigma$ from Eq.(\ref{eq:MFpatternFormation}) for two values of repulsion: $\beta=0.2$ (red) and $\beta=1.0$ (blue). Plot markers correspond to $p=0.1$ (circles), $p=0.4$ (squares),  $p=0.7$ (x's), and $p=1.0$ (triangles). }
\label{fig4}
\end{figure}

In Fig.\ref{fig4} we show that the violation of $\text{MSE}\!\rightarrow\!0$ as $\sigma\!\rightarrow\!0$
is a deterministic nonlinear effect, which is apparent when $\beta$ is large. To demonstrate, we consider target densities of the form $\overline{y}_{i}\!=\!p\cdot y_{i}^{*} + (1-p)/M$, where $p\in[0,1]$ and $\{y_{i}^{*}\}$ is a target with significant spatial fluctuation, $\epsilon \lesssim 1$. In effect, the parameter $p$ interpolates between uniform and spatially complex targets. Two examples are shown in Fig.\ref{fig4} (a); the top corresponds to a target with $p\!=\!0.2$, while the bottom has $p\!=\!1.0$. In panel (b) we vary $p$ and plot the steady-state $\text{MSE}$ from Eq.(\ref{eq:MFpatternFormation}) as a function of $\sigma$ for two values of repulsion: $\beta=0.2$ (red) and $\beta=1.0$ (blue). The different plot markers signify $p=0.1$ (circles), $p=0.4$ (squares),  $p=0.7$ (x's), and $p=1.0$ (triangles). For this target, the graph is a periodic square lattice with $M\!=\!40^{2}$ and $k\!=\!4$. The SF predictions are plotted for each series with dashed and solid lines from Eq.(\ref{eq:SFsteady}), respectively.

For the series with small repulsion in red, all values of $p$ closely track Eq.(\ref{eq:SFsteady}), with the swarm able to 
reach normalized errors of $10^{-6}$ for
$M\sigma\sim\mathcal{O}(10^{-3})$. In this case, even though the targets can be spatially complex, repulsion does not cause a significant deviation from the SF theory. On the other hand, for the series with large repulsion in blue, only the targets with relatively small spatial fluctuations $p=0.1$ and $p=0.4$ track Eq. (\ref{eq:SFsteady}). For $p=0.7$ and $p=1.0$
the $\text{MSE}$ saturates at an error that does not decrease even as the swarm is able to make more precise measurements, $M\sigma\!\rightarrow\!0$. 

Hence, the combination of repulsion and spatial complexity of a target can prevent a swarm from reaching a desired pattern. Since $\beta\!=\!\gamma wN\!/(MV)$, this underscores the importance of choosing a target density  appropriately in order to achieve a faithful realization by the swarm. If the desired resolution is too high (i.e. $V$ is too small) for a given agent size $w$, the agents will experience frequent collisions, and the $\text{MSE}$ will remain high. A similar situation occurs if $N$ is too high for fixed $V$. The ejection rate $\gamma$ can be used to characterize the willingness of agents to remain in overlapping positions. However, this value is usually dictated by the physical characteristics of the agents themselves, and cannot be tuned to suit a given target. It is therefore important to balance the size of the target, the number of agents, and the size of the patches that a target is partitioned into.
%\textcolor{black}{Add a few wrap-up remarks...}
%\textcolor{black}{Limiting-case analytical demonstration?}

\section{\label{sec:D} DISCUSSION} %AND CONCLUSION}
Using swarms of simple mobile agents to collectively solve a prescribed task in a way that is self-organized, robust to perturbation, and without constant external control, is an area of great interest. In this work, we developed a stochastic approach wherein a swarm of agents moves to collectively produce a density pattern that is close to a target. Agents within the swarm know the target density-profile, make stochastic and uncertain measurements at their current location and a randomly selected nearby location, and move in such a way as to reduce the error with the target, based on measurement outcomes. By performing a mean-field and large-$N$ analysis we were able to analytically determine: how fast a swarm produces a steady-state pattern, what the steady-state error is compared to the target, what the effect of large but finite numbers of agents is on the error, and identify multiple crossover regimes between behaviors as a function of physical parameters, such as measurement error, numbers of agents, and collision rates.     

Going further, one of the advantages of a statistical physics approach for targeted swarm density formation, is that including more realistic generalizations into the framework and analysis is relatively straightforward. One can imagine many such generalizations, a few of which we discuss here briefly. We point out that by including effects and limiting factors systematically into a physical framework, one can understand how each changes the behavior, piece by piece.  

\textcolor{black}{First, we assumed that effectively all agents know the target density to within stationary Gaussian uncertainty.}  
%or can accurately measure the target density (within a neighborhood of their own position); 
The difficulty for the swarm was in collectively producing the target given random and uncoordinated swarm-density measurements with error, stochasticity in the motion, repulsion, finite numbers of agents, etc. This assumption could be relaxed in several ways. One way is to assume that each agent is responsible for a subdomain of the total. Within the subdomain an agent knows the target, and has restricted motion in the manner of mobility\cite{PhysRevX.1.011001}. Another way is for the target to be known imperfectly by a small number of leader agents, which communicate to the rest of the swarm and concurrently form a consensus among themselves as the swarm evolves\cite{4542870}. Yet another way is to have physical markers that are embedded in the environment, which for instance, alter the motion of agents through the domain, similar to stigmergy in biological systems\cite{HEYLIGHEN20164}.

Another important aspect involves measurement uncertainty for the swarm density. \textcolor{black}{In this work we analyzed in detail the case where measurements had Gaussian uncertainty with {\it constant} covariance.} However, in general, one would expect local patch measurements to have error that is some function of the true densities: increasing or decreasing with the density depending on the physical process of measurement. \textcolor{black}{For instance, in biological systems, density can be sensed indirectly through chemical concentrations, in which case the sensing error tends to decrease when more agents are present\cite{Camley_2018}. On the other hand, for robotic swarms with visual sensing, for example, the opposite is expected.} \textcolor{black}{As a straightforward extension of the analysis presented, one can study the case where measurement error is Gaussian but with covariance that is an explicit function of the local swarm density. Interestingly, it is simple to show that the small-fluctuation analysis presented in Sec.\ref{sec:SF} is unchanged. Namely, density-dependent uncertainty appears at second order in Eq.(\ref{eq:FirstOrderKreg}). Therefore, qualitatively, density-dependent uncertainty produces swarm-density error that is similar to Fig.\ref{fig4}, where deviation from the small-fluctuation analysis becomes relevant only for sharp target densities with small measurement uncertainty.}

%In an experimental setting, for example with mobile robots, one would calibrate and fit the cumulative distribution function for sensing and measurement, put the model into Eq.(\ref{eq:move}), and analyze the result.

\textcolor{black}{In addition to target and density error, positional error is another problem that swarming agents must overcome. In a robotics context, positional uncertainty can be minimized through additional localization dynamics\cite{ULLAH2024100651}
including recursive Bayesian methods and global localization from an overhead system\cite{7370454}. For biological systems, the position itself must be sensed (directly or indirectly) by agents\cite{PhysRevE.105.044410}. In the latter context, positional uncertainty can be reduced by aggregating estimates from multiple agents organized into spatial clusters, or with the addition of specialized cells that are capable of sensing position more accurately\cite{PhysRevE.100.032417}.
Generically, if position estimates are obtained from a series of independent processes, we expect the uncertainty to be normally distributed, and thus contribute an additional source of variance that can be added to the target and swarm-density uncertainties addressed in this work.} 

%This is an important topic, to be sure. But, it is best addressed in a separate work.

%\textcolor{black}{Localization and position error... instead of the current paragraph?}

%In addition, there are other generalizations having to do with the local dynamical rules for the agents, which could produce interesting changes in the swarm pattern formation. A straightforward generalization would be to study swarms that seek to minimize other error metrics with respect to the target, beyond for instance, the squared error studied. Moreover, other cooperative rules and interactions could be included, whereby, for example, groups of agents coordinate measurements with their local neighbors and move together as a group to reduce the error with the target.  

\textcolor{black}{Another complication 
pertains to the fact that the number of agents in a swarm may vary over time scales that are relevant for the pattern-formation dynamics, which is particularly important for small biological systems. First, to accommodate number fluctuations, one can add stochastic reactions representing birth and death processes to the dynamics\cite{alma991007910019706535}, similar to Eqs.(\ref{eq:move}-\ref{eq:repul}). In this case, however, the state space dynamics should be analyzed in terms of the absolute number of agents at each patch, e.g., $n_{i}(t)$ at patch $i$ and time $t$. Second, if a swarming agent targets a certain absolute number profile $\{\overline{n}_{1},\overline{n}_{2},...,\overline{n}_{M}\}$,
and it measures the number of agents at its local patch $i$ and 
a neighboring patch $j$, $\hat{n}_{i}$ and $\hat{n}_{j}$, respectively, then the sensing-based decision variable $\hat{z}_{ij}$ used throughout, instead becomes $\hat{z}_{ij}\!=\!\overline{n}_{i}-\hat{n}_{i}-\overline{n}_{j}+\hat{n}_{j}$. Once updated, one can analyze the swarm dynamics of the absolute number field in a manner that is similar to what is done in the present work for density, as a function of the birth and death dynamics. For instance, if the time scale for observing a significant fluctuation in the total number of agents is long compared to $1/|\lambda_{M-1}|$ from Eqs.(\ref{eq:LinearSolution1}), we expect the current theory to provide an accurate approximation.}

Even given our current assumptions, however, open theoretical  questions remain such as the uniqueness of the swarm steady-state from arbitrary initial conditions and the effect of finite-size fluctuations in the small-$N$ regime. Both the generalizations and open-questions discussed will be addressed in future work. Nevertheless, we take an important first step, from a theory perspective, in understanding swarms of simple mobile agents that autonomously produce a desired pattern in space from a statistical physics perspective, which answers basic questions about the swarm behavior as a function of physical processes and parameters.

\textcolor{black}{Going beyond theory, we note that our work lends itself to experimental realization and testing as well. 
For example, in terms of robotics experiments, one could use ground robots for the mobile agents, which are equipped with visual, acoustic, or RF sensing capabilities to implement the target-density formation. Depending on the sensing modality, one would need to calibrate how the sensed signal from a robot corresponds to a local number of agents within a region. For acoustic sensing, for instance, the local intensity measured by a robot is expected to be proportional to the number of nearby robots, assuming that each is a random source with approximately equal amplitudes. In the visual case, one might use a machine-learning approach, and train a regression model to predict the number or density of robots in a patch from a camera image (or series of images) as collected by a robot\cite{10.1007/978-981-97-4152-6_17}. Consistent with our framework, in the experiment, the robots would have access to the target density and the pre-trained calibration model, and then move through patches in a domain according to the sensing-based rules that we laid out in Sec.\ref{sec:SSD}. In addition, repulsion would have to be calibrated as well based on, for example, the collision-avoidance capabilities of the particular robots. Experimental testing of this sort is an important next step for future work.} 

%Second, Other distance metrics, cooperative dynamics, and sensing and motion network are different. 
%Fourth localization 

%Neverthless, we have.... summary. 
%Comment on the emergence and stability. The number of parameters is given by the target and physics, but it does not have any scaling with the number of agents. And we don't prepgram 

%\begin{itemize}
%\item Things left open: uniqueness?  
%\item Small $N$
%\item List (some of the) many generalizations one can study in a similar way 
%\end{itemize}

%\section{\label{sec:C} CONCLUSIONS}
%\begin{itemize}
%\item Summary and next steps
%\end{itemize}\\

%\section*{ACKNOWLEDGEMENTS}
\section{ACKNOWLEDGEMENTS}
JH and IBS were supported by the U.S. Naval Research Laboratory funding
(N0001424WX00021), and the Office of Naval Research (N0001425GI01182) and (N0001425GI01158). GS was supported by the Office of Naval Research (N0001425GI01182).
KSK was supported by the Office of Naval Research (N00014-23-1-2434). 

\begin{appendix}
\section{\label{sec:A}APPENDIX}
In this appendix, we will show that the small fluctuation results presented are similar for general networks with binary and symmetric adjacency matrices. As in Sec.\ref{sec:SF}, we are interested in local convergence of a swarm to a target density that is near the random-walk steady state\cite{MASUDA20171} of the patch domain network -- the steady state formed when the sensing uncertainty $\sigma$ becomes arbitrarily large. Namely, we write the target density in the form
\begin{equation}
\label{eq:Bexp}
\overline{y}_{i}=\frac{k_{i}}{\left< k\right>\!M}\big(1+\epsilon \;\overline{\!\!f}_{i}\big), 
\end{equation}
where we define the perturbation amplitude
\begin{equation} 
\epsilon^{2}=\sum_{i}\Big(\frac{k_{i}\overline{y}_{i}}{\left< k\right>\!M}-1\Big)^{2}\!/M. 
\end{equation}
When $\epsilon\!\ll\!1$, we expect the swarm density to take the power series form 
\begin{equation}
\label{eq:Aexp}
y_{i}(t)=\frac{k_{i}}{\left< k\right>\!M}\big(1+\epsilon f_{i,1}+\epsilon^{2}f_{i,2}+...\big). 
\end{equation}

Next, we substitute Eq.(\ref{eq:Bexp}) and Eq.(\ref{eq:Aexp}), into Eq.(\ref{eq:MFpatternFormation}) and collect powers in $\epsilon$. In order to isolate the effects of network topology, we let $\beta\!=\!0$ for this analysis. At $\mathcal{O}(\epsilon)$ the result is
%\begin{align} 
%&\frac{2}{\alpha}D\dot{f_{1}}=L\Big( I+\frac{4}{\sqrt{2\pi}M\sigma\left< k\right>}D\Big)f_{1} \nonumber \\
%&\;\;\;\;\;\;\;\;\;\;\;\;\;-\frac{4}{\sqrt{2\pi}M\sigma\left< k\right>}LD f^{T}, 
%\end{align}
%\begin{align} 
%\label{eq:Alin1}
%&\frac{2}{\alpha}\dot{x}=L\Bigg(D^{-1}+\frac{4}%{\sqrt{2\pi}M\sigma\left< k\right>}I\Bigg)x \nonumber %\\
%&\;\;\;\;\;\;\;\;\;\;\;\;\;-\frac{4}%{\sqrt{2\pi}M\sigma\left< k\right>}LD f^{T}, 
%\end{align}
\begin{align} 
\label{eq:Alin1}
&\frac{2}{\alpha}\dfrac{dx}{dt}=L\Bigg(D^{-1}+\frac{4}{\sqrt{2\pi}M\sigma\left< k\right>}I\Bigg)x \nonumber \\
&\;\;\;\;\;\;\;\;\;\;\;\;\;-\frac{4}{\sqrt{2\pi}M\sigma\left< k\right>}LD\;\overline{\!\!f}, 
\end{align}
where $L$ is the network Laplacian, $D$ is the diagonal degree matrix, $I$ is the identity matrix, and $x\!\equiv\!D f_{1}$.

The solution of Eq.(\ref{eq:Alin1}), depends on the properties of the matrix
\begin{align} 
Q=L\Bigg(D^{-1}+\frac{4}{\sqrt{2\pi}M\sigma\left< k\right>}I\Bigg),   
\end{align}
which we need to establish. First, let us write $Q\!=\!LR$, where $R$ is the positive diagonal matrix $R_{ij}=\delta_{ij}(r+k_{j}^{-1})$ and $r\!=\!4\big/[\sqrt{2\pi}M\sigma\left< k \right>]$. Similar to the Laplacian, $Q$ has column-sums that are equal to zero, $\sum_{i}Q_{ij}\!=\!\sum_{i}L_{ij}(r+k_{j}^{-1})\!=\!(r+k_{j}^{-1})\sum_{i}L_{ij}\!=\!0$. Therefore, if we denote the eigenvalues $\{\lambda_{q}\}$ and right eigenvectors $\{\Lambda_{q}\}$ of $Q$, then for $\lambda_{q}\!\neq\!0$, $\sum_{i}\Lambda_{i,q}\!=\!0$. In addition, $Q$ inherits the single zero eigenvalue of $L$, given that the network is connected. To see this, we note that $Lv_{M}\!=\!0$, where $v_{M}$ is the homogeneous mode of the Laplacian mentioned in the main text. Substituting $L\!=\!QR^{-1}$ into $Lv_{M}\!=\!0$, we get the zero-eigenvalue equation $Q\Lambda_{M}\!=\!0$ with $\Lambda_{M}\!\sim\!R^{-1}v_{M}$. The last property, which we want to show is that all nonzero eigenvalues of $Q$ are negative, just as for $L$. The first step is to note that $Q$ has the same eigenvalues as $R^{1/2}LR^{1/2}$:
%$\det(\lambda  I-Q)=\det(\lambda  I-LR)=\det(\lambda R^{-1/2}R^{1/2}-R^{-1/2}R^{1/2}LR^{1/2}R^{1/2})=\det(R^{-1/2})\det(\lambda  I-R^{1/2}LR^{1/2})\det(R^{1/2})=\det(\lambda  I-R^{1/2}LR^{1/2})=0$, 
\begin{align}
&\det(\lambda  I-Q)=\det(\lambda  I-LR)= \nonumber \\
&\det(\lambda R^{-1/2}R^{1/2}-R^{-1/2}R^{1/2}LR^{1/2}R^{1/2})= \nonumber \\
&\det(R^{-1/2})\det(\lambda  I-R^{1/2}LR^{1/2})\det(R^{1/2})=\nonumber \\
&\det(\lambda  I-R^{1/2}LR^{1/2})=0,
\end{align}
since $R^{-1/2}$ is non-singular. Next, we consider $w^{\intercal}R^{1/2}LR^{1/2}w$, where $w$ is an arbitrary vector. %and $\intercal$ denotes the transpose operation
Since $R^{1/2}$ is non-singular, we can write $d=R^{1/2}w\!\neq\!0$, from which we find $w^{\intercal}R^{1/2}LR^{1/2}w=d^{\intercal}Ld$, using $d^{\intercal}=(R^{1/2}w)^{\intercal}=w^{\intercal}(R^{1/2})^{\intercal}$ and $R^{1/2}=(R^{1/2})^{\intercal}$. Since $d^{\intercal}Ld\leq0$, 
$Q$ inherits the negative semi-definite property of $L$ \cite{mieghem_2010}. 

%\par
%\textcolor{blue}{To prove that $Q = LR$ is a negative semi-definite matrix, note that $R$ is positive definite symmetric, and therefore can be written as the square of a symmetric matrix, namely $R = BB$, such that $B= B^\intercal$. That allows us to express $Q$ as
%\begin{equation*}
%Q = LBB = B^{-1}(BLB)B
%\end{equation*} 
%from which it follows that $Q$ is similar to $BLB$, and hence has identical spectrum. Next we show that $BLB$ is negative definite.}
%\par
%\textcolor{blue}{
%To see this, we need to show that for any non-zero vector $v \in \mathbb{R}^M$ the quadratic form $v^\intercal (BLB) v \leq 0$. Due to the symmetry $B$, we have: 
%\[
%v^\intercal (BLB) v = (vB)^\intercal L (Bv) = w^\intercal L w 
%\]
%where $w = Bv \neq 0$ if and only if $v \neq 0$ because $B$ is also nonsingular. But $L$ is negative semi-definite by definition and therefore $w^\intercal L w \leq 0$ for any nonzero $w \in \mathbb{R}^M$, which completes the proof [GS]
%}

Now, we can use the properties of $Q$ and solve for the local linear dynamics. First, because of normalization $\sum_{i}y_{i}(t)\!=\!1$, we have $\sum_{i}x_{i}(t)\!=\!0$. If we write $x_{i}(t)\!=\!\sum_{q}h_{q}(t)\Lambda_{i,q}$, normalization implies $\sum_{q}h_{q}(t)\sum_{i}\Lambda_{i,q}\!=\!0$. Recalling that only the zero-mode has non-zero vector sum, $\sum_{i}x_{i}(t)\!=\!h_{M}(t)\sum_{i}\Lambda_{i,M}=0$, and therefore  $h_{M}(t)\!=\!0$. Namely, as with $k$-regular networks, we can simply ignore the zero-mode. 

For the other modes, we substitute $x_{i}(t)\!=\!\sum_{q}h_{q}(t)\Lambda_{i,q}$ into Eq.(\ref{eq:Alin1}), and take the inner product with the left (row) eigenvectors of $Q$, which we denote $\{W_{q}\}$. For the $q$th mode the result is
\begin{align} 
\label{eq:Alin2}
%\frac{2}{\alpha}\dot{h}_{q}=\lambda_{q}h_{q}-\frac{4}{\sqrt{2\pi}M\sigma\left< k\right>}W_{q}LDf^{T}.
\frac{2}{\alpha}\dfrac{d h_{q}}{\!\!dt}=\lambda_{q}h_{q}-\frac{4}{\sqrt{2\pi}M\sigma\left< k\right>}W_{q}LD\;\overline{\!\!f}.
\end{align}
The solution to Eq.(\ref{eq:Alin2}) is
\begin{align} 
h_{q}(t)=\big(h_{q}(t\!=\!0)-H_{q}\big)e^{\alpha\lambda_{q}t/2}+H_{q},
\end{align} 
where 
\begin{align}
H_{q}=\frac{4}{\sqrt{2\pi}M\sigma\left< k\right>\lambda_{q}}W_{q}LD\;\overline{\!\!f}.   
\end{align}

Given that $Q$'s eigenvalues are all negative, except for excluded single zero eigenvalue, we have a monotonic decay to a unique steady-state, $h_{q}(t\rightarrow\infty)\!=\!H_{q}$. Hence, altogether, we have shown that the swarm pattern formation over a domain with symmetric networks is also locally convergent, similar to the $k$-regular examples treated in the main text. However, the error with respect to the target density depends on network properties in a more complicated way, in addition to physical parameters. An interesting avenue for future work would be to derive optimal networks for a given target density, given the linear approximation presented, for example.     
\end{appendix}

%\setkeys{DensityFormation2}{articletitle = true}

%\bibliographystyle{unsrt}
%\bibliographystyle{apalike}
%\bibliographystyle{acm}

%\bibliographystyle{ieeetr}
%\bibliographystyle{IEEEtran}
%\bibliographystyle{apssamp}

%\bibliographystyle{apsrev4-1}
\bibliography{DensityFormation2}

\end{document}